\documentclass[english]{article}
\usepackage[T1]{fontenc}
\usepackage[latin9]{inputenc}
\usepackage{amsmath}
\usepackage{amssymb}
\usepackage{esint}

\makeatletter
\usepackage{jheppub}

\newcommand{\tm}{{\tilde{m}}}
\newcommand{\tn}{{\tilde{n}}}

\newcommand{\tf}{{\tilde{f}}}
\newcommand{\dirac}{\nabla \!\!\!\!\!/\;}
\renewcommand{\[}{\begin{equation}}
\renewcommand{\]}{\end{equation}} 
\newcommand{\ZZ}{{\mathbb Z}}
\newcommand{\RR}{{\mathbb R}}
\newcommand{\ra}{\rightarrow}
\author[a]{Anton Kapustin,} 
\author[b]{Brian Willett} 
\author[c]{and Itamar Yaakov} 
\affiliation[a]{Department of Physics, California Institute of Technology,\\ 1200 E. California Blvd, Pasadena, CA} 
\affiliation[b]{Institute for Advanced Study, Einstein Drive, Princeton, NJ} 
\affiliation[c]{Department of Physics, Princeton Univerity, Princeton, NJ} 
\emailAdd{kapustin@theory.caltech.edu} 
\emailAdd{bwillett@ias.edu} 
\emailAdd{iyaakov@princeton.edu}

\usepackage{babel}

\makeatother

\usepackage{babel}
\begin{document}

\title{Exact results for supersymmetric abelian vortex loops in $2+1$ dimensions}

\author{\abstract{We define a class of supersymmetric defect loop operators
in $\mathcal{N}=2$ gauge theories in $2+1$ dimensions. We give a
prescription for computing the expectation value of such operators
in a generic $\mathcal{N}=2$ theory on the three-sphere using localization.
We elucidate the role of defect loop operators in IR dualities of
supersymmetric gauge theories, and write down their transformation
properties under the $SL(2,\mathbb{Z})$ action on conformal theories
with abelian global symmetries.}  }

\maketitle
\tableofcontents{}

\section{Introduction}

Quantum field theories admit a variety of operators defined not by
insertions of the fundamental fields, but by constraints which change
the domain of the path integral in field space. An operator defined
by such a prescription is called a defect operator. A famous example
is the twist operators of 2d conformal field theory. More generally,
one can define a defect operator inserted along a submanifold $L$
by deleting $L$ and requiring the fields to have prescribed singularities
as one approaches $L$. The effect of the insertion can be ``measured''
by evaluating the expectation values of ordinary operators. When one
can detect the presence of the defect from afar (for example, because
some field strengths are now required to belong to a nontrivial cohomology
class), the insertion is said to have created topological disorder.


The first example of a defect operator in gauge theory is probably
the 't Hooft loop operator in 4d gauge theories \cite{'tHooft:1977hy}
which can be used as an order parameter for Higgs phases. It also
plays an important role in the context of electric-magnetic duality
of $\mathcal{N}=4$ gauge theories in $3+1$ dimensions \cite{Montonen:1977sn,Witten:1978mh,Kapustin:2005py}.
The duality exchanges states with electric and magnetic charge and
therefore exchanges Wilson loop operators and 't Hooft loop operators.

There is a somewhat similar story for 3d gauge theories. In such theories
there is a often a duality which exchanges elementary excitations
in the Coulomb phase with Abrikosov-Nielsen-Olesen vortices in the
Higgs phase. A 3d operator creating a very heavy vortex with world
line $L$ is a defect loop operator and may be regarded as analogous
to the 't Hooft loop operator in 4d creating a monopole. Such a defect
operator is defined by the fact that the gauge field has a fixed holonomy
around any small loop linking $L$.  These operators were studied in \cite{Drukker:2008jm}.  In 4d an analogous construction
gives a surface defect \cite{Gukov:2006jk}.

It should be noted that the definition of the vortex loop operator
is independent of the existence of the Higgs phase, or vortex solutions,
or even of a dynamical gauge field. The definition makes perfect sense
in the topological pure Chern-Simons theory as a defect in the gauge
connection, although the defect, in that case, can be identified with
a Wilson loop \cite{Moore:1989yh,Witten:1988hf}. In a theory possessing
an abelian global symmetry, a vortex loop can be defined by gauging
a global symmetry using a non-dynamical flat connection satisfying
the holonomy condition. It is also not necessary that the defect be
defined on a closed loop, however, to preserve gauge invariance, an
open contour must extent to the boundary of spacetime.

Much more can be said about defect operators when the theory is supersymmetric.
All of the operators mentioned above have BPS analogues in supersymmetric
theories in $2+1$ and in $3+1$ dimensions. Extending the definition
of a defect so as to preserve a fraction of the supersymmetry can
require imposing conditions on additional fields. This is analogous
to the inclusion of fields other than the connection in the definition
of a supersymmetric Wilson loop. In this work, we will define the
supersymmetric analogue of the vortex loop. Exact computation of the
expectation value for a supersymmetric defect may be feasible by employing
localization techniques (\cite{Pestun:2007rz,Blau:1995rs,Witten:1988ze}).
This was carried out for the supersymmetric version of the 't Hooft
loop on $\mathbb{S}^{4}$ in \cite{Gomis:2011pf}. Here, we extend
previous results for localization of supersymmetric gauge theories
in $2+1$ dimensions (see \cite{Kapustin:2009kz} for the original
derivation and \cite{Marino:2011nm} for a review) to include the
supersymmetric vortex loop. We will also discuss the role played by
supersymmetric vortex loops in the context of mirror symmetry.

In section \ref{sec:Definition-of-the-abelian-defect}, we define
several versions of the abelian vortex loop. We discuss the operator's
transformation under Witten's $SL(2,\mathbb{Z})$ action on conformal
field theories. We then extend the definition to accommodate supersymmetry.
In section \ref{sec:Localization-of-the-supersymmetric-defect-operator},
we employ localization to evaluate the expectation value of the supersymmetric
vortex loop for a generic superconformal $\mathcal{N}=2$ gauge theory
on $\mathbb{S}^{3}$. The result can be inferred from the $SL(2,\mathbb{Z})$
action. We provide an independent derivation using the original definition.
In section \ref{sec:Duality-with-defect-operators}, we demonstrate,
with a few examples, the role of the vortex loop in IR duality of
$\mathcal{N}=2$ gauge theories. We conclude with a discussion of
possible extensions.

\section{\label{sec:Definition-of-the-abelian-defect}Definition of the abelian
vortex loop}


\subsection{Vortex loop in a gauge theory}

Defect loop operators in 3d gauge theories have been previously introduced
in the context of Chern-Simons theories \cite{Witten:1988hf,Moore:1989yh}.
Such a loop operator is specified by giving a loop $\gamma$ in the
3-manifold $M$ and an element $\beta$ of the Lie algebra $\mathfrak{g}$
of the gauge group $G$. The holonomy around any small loop linking
$\gamma$ of the gauge connection $A$ is required to approach $\beta$
as we shrink the loop size to zero. With this condition, $A$ is singular
on $\gamma$. Separating $A$ into a smooth part $A'$ and a singular
part $A''$, we can write 
\begin{equation}
{F_{A''}}=\beta\star\left[\gamma\right]\label{eq:singular_field_strength}
\end{equation}
where $\star[\gamma]$ is a 2-form current supported on $\gamma$
whose cohomology class is the Poincar� dual of $[\gamma]\in H_{1}(M,\ZZ)$
(we assume that $M$ is orientable). In \cite{Gukov:2006jk}, the
authors defined surface operators in $\mathcal{N}=4$ SYM theory in
$3+1$ dimensions using a similar prescription. The prescription for
a subset of these codimension $2$ operators, the one in which only
the connection is singular, coincides with the definition above (substitute
$\alpha$ for $\beta$). As noted there, the data specifying the singularity
is actually only $e^{i\beta}$, and equation \ref{eq:singular_field_strength}
should be handled with care. If the gauge group $G$ is $U(1)$, $\beta$
is simply a real number. To avoid ambiguity it is sufficient to restrict
the range of $\beta$ to the interval $(-\pi,\pi)$. In what follows,
we will sometimes assume this restriction. We will also set 
\[
q=\frac{\beta}{2 \pi},\qquad q\in\left(-\frac{{1}}{2},\frac{1}{2}\right)
\]

From now on we will assume that $G$ is abelian and will call such
a loop operator a gauge vortex loop.

\subsection{Global vortex loops and the $SL(2,\ZZ)$ action}

We can define a similar loop operator if $G$ is a global symmetry
group rather than a gauge symmetry group. To this end we merely set
the smooth part $A'$ of the gauge field to zero and set the singular
part ($A''$ above) to be a fixed flat connection whose holonomy along
a loop linking $\gamma$ is $e^{i\beta}$. We will call such a loop
operator a global vortex loop. It appears naturally when we consider
the action of Witten's $SL(2,\ZZ)$ on Wilson loops.

Following \cite{Witten:2003ya}, we consider a conformal field theory
in $2+1$ dimension with a choice of an abelian global symmetry current
$J$. We couple $J$ to a background gauge field $A$, and consider
the partition function as a functional of $A$

\[
Z[A]=\int\mathcal{D}\Phi e^{iS[\Phi]+i\int\sqrt{g}d^{3}xJ^{\mu}A_{\mu}+...}
\]
where ``$...$'' refers to seagull terms necessary to ensure invariance
under gauge transformations of $A$. Here $\Phi$ and $S[\Phi]$ are
short hand for the fields and action of the theory. In addition to
the seagull terms, one can add extra terms which are gauge-invariant
functionals of $A$ alone. If we wish to preserve conformal symmetry,
a natural choice is the abelian Chern-Simons term

\[
\frac{i\alpha}{4\pi}\int A\wedge dA
\]
If we want to associate $A$ to a non-trivial principal bundle, then
this is only defined for integer $\alpha$. However, for now, we will
assume that $A$ is a connection on a trivial bundle, and allow arbitrary
real values of $\alpha$. To the triplet $(S[\Phi],J,\alpha)$ we
associate the following functional of $A$:

\[
Z_{J,\alpha}[A]=\int\mathcal{D}\Phi e^{iS[\Phi]+i\int\sqrt{g}d^{3}xJ^{\mu}A_{\mu}+...+\frac{i\alpha}{4\pi}\int A\wedge dA}
\]

Witten defined an $SL(2,\ZZ)$ action on such triplets \cite{Witten:2003ya}.
The action of the $T$ generator is merely a shift of $\alpha$ by
$1$. The action of $S$ is defined by first promoting $A$ to a dynamical
gauge field and then replacing $J$ with the topological $U(1)$ current
for this new gauge field 
\[
J_{top}=\frac{1}{2\pi}\star dA
\]
Together, these operations generate an action of $SL(2,\mathbb{Z})$
on the set of triples. At the level of partition functions, these
operations act as follows:

\[
(T\cdot Z_{J,\alpha})[A]=Z_{J,\alpha+1}[A]
\]

\[
(S\cdot Z_{J,\alpha})[A]=\int\mathcal{D}A'\mathcal{D}\Phi e^{iS[\Phi]+i\int\sqrt{g}d^{3}xJ^{\mu}A'_{\mu}+...+\frac{i\alpha}{4\pi}\int A'\wedge dA'+\frac{i}{2\pi}\int A\wedge dA'}
\]

Our first goal is to extend these operations to observables more general
than the partition function $Z_{J,\alpha}$. Specifically, we would
like to consider insertions of Wilson loops for the background gauge
field $A$. First, it is useful to define a slight generalization
of an abelian Wilson loop. Recall that the abelian Wilson loop operator
is specified by a loop $\gamma:S^{1}\ra M$ and a charge $q\in\RR$
and is defined as an insertion of

\[
e^{iq\int_{\gamma}A}
\]
into the path integral. This can be rewritten as

\[
e^{\frac{i}{2\pi}\int\omega\wedge A}
\]
where $\omega$ is a closed $2$-form current with support along $\gamma$,
defined so that the integrals in the previous two expressions agree
for all $1$-forms $A$. In local coordinates $(r,\theta,z)$ where
the loop lies along the $z$ axis, we can write:

\[
\omega=2\pi q\delta_{\gamma}\equiv q\delta(r)dr\wedge d\theta
\]
Such a term can simply be added to the action, since in the abelian
case we do not have to worry about path ordering the exponential.

We can consider such an insertion for more general 2-form currents
$\omega$. Invariance with respect to infinitesimal gauge transformations
forces $\omega$ to be closed. Invariance with respect to ``large''
gauge transformations (i.e. gauge transformations which are topologically
nontrivial maps from the abelian Lie group $G$ to $M$) requires
the de Rham cohomology class of $\omega/2\pi$ to be integral. This
means that $\omega$ arises as the field strength of some connection
$A_{\omega}$ on a $U(1)$ bundle, and we can write the Wilson loop
as a $BF$ coupling to this new background gauge field:

\[
e^{\frac{i}{2\pi}\int A\wedge dA_{\omega}}
\]
In the case 
\[
\omega=2\pi q\delta_{\gamma}
\]
where $\delta_{\gamma}$ is a 2-form current supported on $\gamma$
which is Poincar� dual to $\gamma$, the integrality condition on
$\omega$ reduces to the requirement that the class $q[\gamma]\in H_{1}(M,\RR)$
is integral. In particular, if $\gamma$ is homologically trivial,
there are no integrality constraints on $q$. Near $\gamma$ the connection
$A_{\omega}$ in suitable coordinates looks as follows: 
\[
A_{\omega}=qd\theta
\]
More generally, if we take $\omega$ to be supported in a small tubular
neighborhood of $\gamma$, we get a regularization of the Wilson loop
along $\gamma$.


With this in mind, we define a insertion of $W_{\omega}$ by 
\[
(W_{\omega}\cdot Z_{J,\alpha})[A]:=\int\mathcal{D}\Phi e^{iS[\Phi]+i\int\sqrt{g}d^{3}xJ^{\mu}A_{\mu}+...+\frac{i\alpha}{4\pi}\int A\wedge dA+\frac{i}{2\pi}\int A_{\omega}\wedge dA}
\]
Let us see how the $SL(2,\mathbb{Z})$ generators act on it. It is
clear that $T$ commutes with $W_{\omega}$, so we only need to consider

\begin{equation}
((S^{-1}W_{\omega}S)\cdot Z_{J,\alpha})[A]=\int\mathcal{D}A_{1}\mathcal{D}A_{2}\mathcal{D}\Phi\exp\bigg(iS[\Phi]+i\int\sqrt{g}d^{3}xJ^{\mu}{A_{1}}_{\mu}+...+\frac{i\alpha}{4\pi}\int{A_{1}}\wedge d{A_{1}}+\label{eq:S_transformation}
\end{equation}

\[
+\frac{i}{2\pi}\int A_{2}\wedge dA_{1}+\frac{i}{2\pi}\int A_{\omega}\wedge dA_{2}-\frac{i}{2\pi}\int A\wedge dA_{2}\bigg)
\]
We can see $A_{2}$ enters only via a term $\int(A_{1}-A_{\omega}-A)\wedge dA_{2}$,
and so the integral over $A_{2}$ produces exactly a delta function
setting $A_{1}=A_{\omega}+A$ \cite{Witten:2003ya}. This leaves

\[
\int\mathcal{D}\Phi\exp\bigg(iS[\Phi]+i\int\sqrt{g}d^{3}xJ^{\mu}({A_{\omega}}_{\mu}+A_{\mu})+...+\frac{i\alpha}{4\pi}\int(A+A_{\omega})\wedge d(A+A_{\omega})\bigg)
\]

\[
=Z_{J,\alpha}[A+A_{\omega}]
\]

Note that, even at $A=0$, this gives an insertion of an operator:

\[
\exp\bigg(i\int\sqrt{g}d^{3}xJ^{\mu}{A_{\omega}}_{\mu}+...+\frac{i\alpha}{4\pi}\int A_{\omega}\wedge\omega\bigg)
\]
In the case where $\omega=2\pi q\delta_{\gamma}$, $A_{\omega}$ is
a flat gauge field with a holonomy $e^{2\pi iq}$, and this operation
has the same effect on the path integral as prescribing that all fields
charged under the current $J$ pick up a fixed monodromy around the
loop $\gamma$. There is also an additional $\alpha$-dependent phase
factor related to the self-linking number of the loop (which must
be regularized by specifying a framing). We see that in this special
case the operation $S$ maps the charge-$q$ Wilson loop for the background
gauge field $A$ to the global vortex loop with holonomy $e^{2\pi iq}$.
More generally, for arbitrary $\omega$ satisfying the above integrality
conditions we can define an operation 
\[
(D_{\omega}\cdot Z_{J,\alpha}[A])=Z_{J,\alpha}[A+A_{\omega}]
\]
What we have demonstrated is that the operation $S$ maps $W_{\omega}$
to $D_{\omega}$:

\[
S^{-1}W_{\omega}S=D_{\omega}
\]
More precisely, the equality holds up to a phase factor which depends
not only on $\omega$, but also on the Chern-Simons coupling of the
theory on which these operations act.

One can rephrase this result in terms of Wilson loops for dynamical
gauge fields. Gauging a symmetry (without adding a Chern-Simons term)
is the same as applying the operation $S$. The resulting theory has
a new global symmetry $U(1)_{J}$ whose current is $\star dB/2\pi$,
where we denoted by $B$ the dynamical gauge field, to distinguish
it from the background gauge field $A$ which couples to the $U(1)_{J}$
current. Noting that $S^{2}=C$ (the charge conjugation), we get 
\[
SW_{\omega}=D_{-\omega}S.
\]
Applying this to the partition function $Z_{J,\alpha}^{ungaged}[B]$,
we learn that 
\[
<W_{\omega}[B]>=D_{-\omega}\cdot Z_{gauge}[A]=Z_{gauge}[A-A_{\omega}].
\]
In particular, setting $\omega/2\pi$ to be the delta-function supported
on a loop $\gamma$, we see that a global vortex loop for the $U(1)_{J}$
symmetry is nothing more than an ordinary Wilson loop in the underlying
gauge field. Similarly, we find 
\[
<D_{\omega}[B]>=W_{\omega}\cdot Z_{gauge}[A]
\]
which shows that, in the absence of a Chern-Simons term, the gauge
vortex loop by itself is somewhat trivial: it merely modifies the
functional dependence of $Z_{gauge}[A]$ on the background gauge field
$A$ which couples to the $U(1)_{J}$ current.

\subsection{Pure Chern-Simons theory}

Before moving on to the supersymmetric version of the vortex loop,
let us briefly comment on how the gauge vortex loop behaves in pure
bosonic Chern-Simons theory. It was argued in \cite{Moore:1989yh}
that such a defect operator should be equivalent to a Wilson loop.
However, we have seen above that for an abelian gauge group the gauge
vortex loop is somewhat trivial, its only effect being a modification
of the $U(1)_{J}$ current by a $c$-number term supported on the
loop. To see that this agrees with the behavior of the Wilson loops,
we recall the formula for the expectation value of a product of Wilson
loops in $U(1)$ Chern-Simons theory at level $k$ \cite{Witten:1988hf}

\begin{equation}
\langle\prod_{a}\exp(iq_{a}\int_{\gamma_{a}}A)\rangle=\exp\bigg(\frac{2\pi i}{k}\sum_{a,b}q_{a}q_{b}\Phi(\gamma_{a},\gamma_{b})\bigg)\label{eq:abwi}
\end{equation}
where $\Phi(\gamma_{a},\gamma_{b})$ is the linking number of the
loops $a$ and \textbf{$b$}. The latter can be written in terms of
the corresponding gauge fields $A_{a}$, with $dA_{a}=\omega_{a}$
the $2$-form delta function supported on $\gamma_{a}$, as

\[
\Phi(\gamma_{a},\gamma_{b})=\int_{\gamma_{b}}A_{a}=\frac{1}{2\pi}\int A_{a}\wedge\omega_{b}
\]
Letting 
\[
\omega=\frac{1}{k}\sum_{a}q_{a}\omega_{a},
\]
we can rewrite the expectation value as follows: 
\[
\langle\prod_{a}\exp(iq_{a}\int_{\gamma_{a}}A)\rangle=\exp\left(\frac{ik}{2\pi}\int A_{\omega}\wedge dA_{\omega}\right).
\]
Thus the insertion of a collection of Wilson loops is equivalent to
a phase factor which depends on $k$ as well as on a flat connection
$A_{\omega}$. This is compatible with the claim that a collection
of Wilson loops in pure abelian Chern-Simons theory is equivalent
to a gauge vortex loop $D_{\omega}$ for some 2-form current $\omega$,
which in turn is trivial up to a phase.

\subsection{Supersymmetric vortex loops}

We would like to extend the considerations above to the supersymmetric
case. Specifically, we will work with theories with $\mathcal{N}=2$
supersymmetry ($4$ real supercharges). It is convenient to work in
$\mathcal{N}=2$ superspace, with fermionic coordinated $\theta_{\alpha}$
and a superspace derivative $D_{\alpha}$. For the theories of interest,
the dynamical fields can be organized into chiral and vector superfields.
The gauge field is part of a vector multiplet, and all fields in this
multiplet take values in the adjoint representation of the gauge group.
A vector superfield $V$ satisfies $V=V^{\dagger}$ and contains a
vector field $A_{\mu}$, a real scalar $\sigma$, a complex spinor
$\lambda_{\alpha}$, and a real auxiliary scalar $D$. Matter fields
live in chiral multiplets and take values in some representation of
the gauge and flavor groups. A chiral superfield $\Phi$ satisfies
${{\bar{D}}_{\alpha}}\Phi=0$, and contains a complex scalar $\phi$,
a complex two component spinor $\psi_{\alpha}$, and an auxiliary
complex scalar $F$.

We are interested mainly in a class of renormalizable gauge theories
with abelian global symmetries. These are defined by a UV action which
includes a kinetic term for the matter fields of the form

\begin{align}
S_{{\text{charged matter kinetic}}} & =-\int{{d^{3}}x{d^{2}}\theta{d^{2}}\bar{\theta}\sum\limits _{i}{({\Phi_{i}}^{\dag}{e^{2V}}{\Phi_{i}})}}\label{eq:ChargedMatterKinetic}\\
 & =\sum\limits _{i}{\int{{d^{3}}x\left({{{({D_{\mu}}\phi)}_{i}}{{({D^{\mu}}\phi)}^{i}}+i{{\bar{\psi}}_{i}}{\gamma^{\mu}}{D_{\mu}}{\psi^{i}}+{F_{i}}{F^{i}}-{\phi_{i}}{\sigma^{2}}{\phi^{i}}+{\phi_{i}}D{\phi^{i}}-{{\bar{\psi}}_{i}}\sigma{\psi^{i}}+i{\phi_{i}}\bar{\lambda}{\psi^{i}}-i{{\bar{\psi}}_{i}}\lambda{\phi^{i}}}\right)}}
\end{align}
and a supersymmetric Yang-Mills action for the fields in the gauge
multiplet 
\begin{align}
S_{{\text{Yang Mills}}} & =\frac{1}{{g^{2}}}\int{{d^{3}}x{d^{2}}\theta{d^{2}}\bar{\theta}T{r_{f}}\left({\frac{1}{4}{\Sigma^{2}}}\right)}\label{eq:N2YangMills}\\
 & =\frac{1}{{g^{2}}}\int{{d^{3}}xT{r_{f}}\left({\frac{1}{2}{F_{\mu\nu}}{F^{\mu\nu}}+{D_{\mu}}\sigma{D^{\mu}}\sigma+{D^{2}}+i\bar{\lambda}{\gamma^{\mu}}{D_{\mu}}\lambda}\right)}
\end{align}
where $\Sigma$ is a linear multiplet defined by

\[
\Sigma={{\bar{D}}^{\alpha}}{D_{\alpha}}V
\]

\[
{\Sigma^{\dag}}=\Sigma
\]

\[
{D^{\alpha}}{D_{\alpha}}\Sigma={{\bar{D}}^{\alpha}}{{\bar{D}}_{\alpha}}\Sigma=0
\]
In addition, one can allow supersymmetric completions of Chern-Simons
terms. In the abelian case, these have a very simple superspace expression

\begin{align}
S_{{\text{abelian Chern Simons}}} & =\frac{k}{{4\pi}}\int{{d^{3}}x{d^{2}}\theta{d^{2}}\bar{\theta}T{r_{f}}\left({V\Sigma}\right)}\label{eq:N2AbelianChernSimons}\\
 & =\frac{k}{{4\pi}}\int{{d^{3}}x\text{Tr}_{f}\left({{\varepsilon{}^{\mu\nu\rho}}{A_{\mu}}{\partial_{\nu}}{A_{\rho}}-\bar{\lambda}\lambda+2D\sigma}\right)}
\end{align}
As a slight generalization of this, we can also consider a ``off-diagonal''
Chern-Simons term, also known as a $BF$ term, coupling two or more
different abelian gauge fields. The supersymmetric completion of this
has the form

\begin{align}
S_{{\text{BF}}} & =\frac{{k_{ij}}}{{4\pi}}\int{{d^{3}}x{d^{2}}\theta{d^{2}}\bar{\theta}T{r_{f}}\left({{\Sigma^{i}}{V^{j}}}\right)}\label{eq:N2BF}\\
 & =\frac{{k_{ij}}}{{4\pi}}\int{{d^{3}}x\left({{\varepsilon^{\mu\nu\rho}}{A^{j}}_{\mu}{\partial_{\nu}}{A^{i}}_{\rho}-\frac{1}{2}{{\bar{\lambda}}^{j}}{\lambda^{i}}+{D^{i}}{\sigma^{j}}}\right)}
\end{align}
We will consider the $S^{3}$ partition function for these theories
with insertions of Wilson loop and defect loop operators. The supersymmetry
transformations on $S^{3}$ are generated by Killing spinors \cite{Kapustin:2009kz}.
We will need the transformations of the vector multiplet fields under
the $S^{3}$ supersymmetry generated by a particular killing spinor
$\varepsilon$ 
\begin{align}
 & \delta{A_{\mu}}=-\frac{i}{2}{\lambda^{\dag}}{\gamma_{\mu}}\varepsilon\hfill\\
 & \delta\sigma=-\frac{1}{2}{\lambda^{\dag}}\varepsilon\hfill\\
 & \delta D=-\frac{i}{2}{D_{\mu}}{\lambda^{\dag}}{\gamma^{\mu}}\varepsilon+\frac{i}{2}[{\lambda^{\dag}},\sigma]\varepsilon+\frac{1}{4}{\lambda^{\dag}}\varepsilon\hfill\\
 & \delta\lambda=(-\frac{i}{2}{\varepsilon^{\mu\nu\rho}}{F_{\mu\nu}}{\gamma_{\rho}}-D+i{\gamma^{\mu}}{D_{\mu}}\sigma-\sigma)\varepsilon\hfill\label{eq:GauginoVariation}\\
 & \delta\lambda^{\dagger}=0\hfill
\end{align}
The fermionic symmetry generated by $\varepsilon$ will be used in
\ref{sec:Localization-of-the-supersymmetric-defect-operator} to compute
the expectation value of the defect operator on $S^{3}$ using localization. 

We now attempt to generalize the $SL(2,\mathbb{Z})$ action to the
supersymmetric case. To start, it is natural to define $T$ by simply
adding a supersymmetric Chern-Simons term instead of an ordinary one.
For $S$, we should use the supersymmetric version of the $BF$ term
and integrate over the entire background vector multiplet. It is now
a simple exercise to check that the $SL(2,\mathbb{Z})$ relations
remain satisfied for this generalization. We will not use the $(ST)^{3}=C$
property, so we omit a check of that relation, but it will be important
that $S^{2}=C$, so let us sketch the argument. Consider the supersymmetric
version of \ref{eq:S_transformation} defined via the action \ref{eq:N2BF}
where all vector fields have been extended to $\mathcal{N}=2$ vector
multipelts $V,V_{1},V_{2}$ and $V_{\omega}$. When we integrate over
the second gauge field, the $BF$ coupling gives us a delta function
constraint imposing that the first gauge field is the negative of
the background gauge field, as before. In addition, one can see that
the integration over the auxiliary fields in the second vector multiplet
imposes a similar constraint on the auxiliary fields of the first
vector multiplet. Thus we see that the net effect is to set the first
vector multiplet equal to the negative of the background vector multiplet.

Now consider the supersymmetric generalization of the abelian Wilson
loop operator. For a certain class of loops $\gamma$ preserving supersymmetry,
this has the form

\[
\mbox{exp}\bigg(iq\oint_{\gamma}(A-i\sigma d\ell)\bigg)
\]
For example, on $S^{3}$, the loops must be great circles which are
fibers of the Hopf fibration. This operator is then invariant under
the supersymmetry generated by $\varepsilon$. We define an operator
$W_{\gamma}$ inserting a supersymmetric Wilson loop in a background
vector multiplet as before. Finally, we define the supersymmetric
vortex loop by the prescription $D_{\gamma}=S^{-1}W_{\gamma}S$.

As before, we can integrate out the auxiliary vector multiplets $V_{1}$
and $V_{2}$ to obtain a description of the defect terms of the original
fields alone. To start, let us write the part of the action containing
$V_{2}$. In terms of component fields the action (in Euclidean signature)
looks as follows:

\[
S[\Phi,V_{1},V_{2}]=...+\frac{i}{2\pi}\int(-A_{2}\wedge dA_{1})+\frac{i}{2\pi}\int d^{3}x(-\sigma_{2}D_{1}-\sigma_{1}D_{2}+\frac{1}{2}(\lambda_{1}^{\dagger}\lambda_{2}+\lambda_{2}^{\dagger}\lambda_{1}))+iq\int_{\gamma}(A_{2}-i\sigma_{2}d\ell)
\]
For simplicity we set to zero the background vector multiplet. We
see that the integrals over $D_{2}$ and $\lambda_{2}$ set $\sigma_{1}$
and $\lambda_{1}$ to zero, while the integrals over $A_{2}$ and
$\sigma_{2}$ impose the constraints

\[
dA_{1}=2\pi q\delta_{\gamma},\;\;\;\star D_{1}=-2\pi iq\delta_{\gamma}\wedge d\ell.
\]
Here $d\ell$ is the volume 1-form on $\gamma$ and $\delta_{\gamma}$
is the $2$-form Poincar� dual to $[\gamma]$, as before. Note that
$D_{1}$ is purely imaginary, which violates the usual reality condition
on $D$.

As before, we can generalize this operator to account for more general
background gauge multiplet configurations. We to specialize to $S^{3}$,
and pick a supercharge $\delta$ corresponding to the Killing spinor
$\epsilon$. Then the BPS condition for an abelian vector multiplet
is \cite{Kapustin:2009kz}

\[
0=(i\gamma^{\mu}(-\star F_{\mu}+\partial_{\mu}\sigma)-(D+\sigma))\epsilon
\]
We would like to find configurations for which only $F$ and $D$
are non-zero. Using $v^{\mu}\gamma_{\mu}\epsilon=\epsilon$, where
$v^{\mu}$ is the Killing vector along the Hopf fibration as in \cite{Kapustin:2009kz},
we see we can take:

\[
F=2\pi f(x)\star v,\;\;\; D=-2\pi if(x)
\]
for a function $f:S^{3}\rightarrow\mathbb{R}$. The normalization
is for later convenience. Note that the Bianchi identity implies

\[
0=\frac{1}{2\pi}dF=d(f\star v)=df\wedge\star v=\star(v^{\mu}\partial_{\mu}f)
\]
so that $f$ must be constant along the fibers of the Hopf fibration.
Equivalently, we impose that $f$ arises from a function $\tf:S^{2}\rightarrow\mathbb{R}$
by:

\[
f=\tf\circ\pi
\]
where $\pi:S^{3}\rightarrow S^{2}$ is the projection map of the Hopf
fibration. Thus, the operator is really labeled by the function $\tf$
on $S^{2}$. When $\tf$ approaches a delta-function on $S^{2}$,
the corresponding operator approaches the supersymmetric vortex loop
wrapping a fiber of the Hopf fibration.

If we apply this construction to the $U(1)_{J}$ symmetry, we get
an operator which is a natural generalization of the supersymmetric
Wilson loop:

\[
\mathcal{O}_{\tf}=\exp\bigg(i\int_{S^{3}}\sqrt{g}d^{3}xf(x)(v^{\mu}A_{\mu}-i\sigma)\bigg)
\]
That this operator is supersymmetric follows from

\[
\delta(v^{\mu}A_{\mu}-i\sigma)=0
\]
as can be easily checked. The condition $v^{\mu}\partial_{\mu}f=0$
is necessary for gauge-invariance. Taking $\tf$ to be a delta function
on $S^{2}$, so that $f$ is a delta function supported along a great
circle in $S^{3}$, one recovers the ordinary supersymmetric Wilson
loop. Note that we have picked the normalizations so that when $\tilde{f}$
is a delta function on $S^{2}$ integrating to $1$, so that $f$
integrates to $2\pi$ on $S^{3}$, we recover the charge $1$ Wilson
loop.

\section{\label{sec:Localization-of-the-supersymmetric-defect-operator}Localization
in the presence of a vortex loop}

In this section we compute the expectation value of the global vortex
loop on $S^{3}$ by localization. The global symmetry is assumed to
be a $U(1)$ subgroup of the flavor symmetry group. We will present
three approaches to the calculation which yield the same result. Applying
the localization procedure in the presence of the defect requires
some regularization and the agreement of the approaches presented
below gives us confidence in the validity of the computation. We begin
with a quick review of localization for 3d gauge theories. Additional
details can be found in \cite{Kapustin:2009kz}.

\subsection{Localization of 3d gauge theories}

The expectation value of BPS operators in 3d $\mathcal{N}=2$ superconformal
gauge theories can be computed by localization on $S^{3}$. The relevant
result, derived in \cite{Kapustin:2009kz}, is that deformation invariance
allows us to reduce the computation of the infinite dimensional path
integral with BPS operator insertions to a matrix model with the integration
domain given by the Lie algebra of the gauge group. The data entering
the computation is a UV action with gauge group $G$, Lie algebra
$\mathfrak{g}$ and Chern-Simons levels $k_{i}$, a set of representations
$R_{i}$ for the chiral matter multiplets and the IR conformal dimensions
(equivalently R-charges) $\Delta_{i}$ for each chiral multiplet.
The integration measure for the matrix model is then 
\begin{equation}
\frac{1}{{{\text{Vol}}(G)}}da{|_{a\in{\text{Ad}}(\mathfrak{g})}}\label{eq:measure}
\end{equation}
The contribution of a level $k$ Chern-Simons term (for a simple gauge
group factor associated to $a$) is 
\[
{e^{-i\pi k{\text{Tr}}\left({a^{2}}\right)}}
\]
A Fayet-Iliopoulos term with coefficient $\eta$ contributes 
\[
{e^{2\pi i\eta{\text{Tr}}\left(a\right)}}
\]
Every dynamical gauge multiplet contributes 
\[
Z_{{\text{1 - loop}}}^{{\text{gauge multiplet}}}(a)={\text{de}}{{\text{t}}_{{\text{Ad(}}\mathfrak{g})}}\left(2\sinh(\pi a)\right)=\prod\limits _{\rho\in{\text{roots}}(\mathfrak{g})}{2\sinh(\pi\rho(a))}
\]
and every dynamical chiral multiplet contributes 
\begin{equation}
Z_{{\text{1 - loop}}}^{{\text{chiral multiplet}}}(a,\Delta)=\frac{\text{det}\mathcal{O}_{F}}{\sqrt{\text{det}\mathcal{O}_{B}}}=\prod\limits _{\rho\in R}{\exp\left({\ell\left({z(\rho(a),\Delta)}\right)}\right)}\label{eq:1_loop_chiral}
\end{equation}
where $\rho$ are the weights of $R$ and 
\[
\ell(z)=-z\log\left({1-{e^{2\pi iz}}}\right)+\frac{i}{2}\left({\pi{z^{2}}+\frac{1}{\pi}{\text{L}}{{\text{i}}_{2}}\left({e^{2\pi iz}}\right)}\right)-\frac{{i\pi}}{{12}}
\]
\[
z(\rho(a),\Delta)=i\rho(a)-\Delta+1
\]
abelian flavor parameters can be incorporated by shifting $\rho\rightarrow\rho+m$.
The insertion of a supersymmetric Wilson loop in a representation
$R$ gives a factor of 
\[
W(a)=\frac{1}{{{\text{dim}}(R)}}{\text{T}}{{\text{r}}_{R}}\left({e^{2\pi a}}\right)
\]
which for an abelian Wilson loop of charge $q$ reduces to 
\[
e^{2\pi qa}
\]
Integration with the measure \ref{eq:measure} of the product of all
relevant contributions yields the exact expectation value.

In computing \ref{eq:1_loop_chiral} we have implicitly assumed a
standard $\delta$ exact term, and hence standard kinetic operators
($\mathcal{O}_{F},\mathcal{O}_{B}$), for the fields in the chiral
multiplet \cite{Kapustin:2009kz,Jafferis:2010un}. In the presence
of the flavor vortex loop, the kinetic term of a charged chiral multiplet
is altered by a background gauge field created by the loop. The new
term and the revised 1-loop contribution are derived below. The contribution
of the vector multiplet is unaffected because it is not charged under
flavor symmetries.

\subsection{Method $1$: using the $SL(2,\mathbb{Z})$ definition of $D$}

The simplest way to extract the effect of inserting a supersymmetric
defect line operator is by using the definition of the operation $D$
as

\[
D_{q}=S^{-1}W_{q}S
\]
where $D_{q}$ is the vortex loop with holonomy $\exp(2\pi iq)$.
There are no integrality constraints on $q$ because the large circle
on $S^{3}$ is homologically trivial. Since we can perform the operations
on the RHS at the level of the matrix model, it should be possible
to compute the LHS indirectly this way. Explicitly, suppose we compute
the partition function as a holomorphic function of a flavor deformation
$m$ (and possibly other parameters which we suppress):

\[
Z(m)
\]
This is the analogue of $Z[A]$ in section $2$. The operation $S$
then tells us to treat this flavor parameter as a gauge parameter,
and integrate over it, with a coupling to an FI parameter $\eta$:

\[
(S\cdot Z)(\eta)=\int dmZ(m)e^{2\pi i\eta m}
\]
The operation $W_{1}$ tells us to insert a charge-$q$ Wilson loop
in the background field corresponding to $\eta$:

\[
(W_{q}S\cdot Z)(\eta)=e^{2\pi q\eta}\int dmZ(m)e^{2\pi i\eta m}
\]
Finally, $S^{-1}$ tells us to integrate over $\eta$ and insert a
new FI term, which we will denote $m'$, with the opposite sign:

\[
(S^{-1}W_{1}S\cdot Z)(m')=\int e^{-2\pi i\eta m'}e^{2\pi q\eta}\int dmZ(m)e^{2\pi i\eta m}
\]
Now to integrate out the variables $\eta$ and $m$, we simply note
that the integral over $\eta$ imposes a delta function which sets
$m=m'+iq$. Thus we are left with:

\[
(D_{q}\cdot Z)(m)=Z(m+iq)
\]
Indeed, this result can be inferred from the 4d perspective by considering
Wilson and 't Hooft loops ending on a 3d boundary \cite{Dimofte:2011ju,Dimofte:2011py}.
This argument was rather indirect; it also raises the question about
the interpretation of poles in the partition function for special
values of $m+iq$. We now proceed to present two more explicit derivations
of this result.

\subsection{Method $2$: smearing the defect}

We return to the smeared Wilson loop

\[
\mathcal{O}_{f}=\exp\bigg(i\int_{S^{3}}\sqrt{g}d^{3}xf(x)(v^{\mu}A_{\mu}-i\sigma)\bigg)
\]
where $f$ is some real function on $S^{3}$ constant along the fibers
of the Hopf fibration, as above. It is convenient to decompose $f(x)$
as

\[
f(x)=\frac{q}{\pi}+f_{o}(x)
\]
where

\[
\int_{S^{3}}\sqrt{g}d^{3}xf_{o}=0
\]
and $q$ is constant, specifically, $q=\frac{1}{2\pi}\int_{S^{3}}\sqrt{g}d^{3}xf$.
Note that this normalization agrees with the case where $f$ is a
delta function supported on a great circle, since the integral of
$f$ should give $2\pi$ times the charge of the Wilson loop.

For general $f$, we can decompose $\mathcal{O}_{f}=\mathcal{O}_{q}\mathcal{O}_{f_{o}}$,
so it suffices to study them separately. Actually, we will find that
it is only $\mathcal{O}_{q}$ which contributes to the localized path-integral.
Specifically, we claim that the operator $\mathcal{O}_{f_{o}}$ can
be absorbed into a shift of the action by a total $\delta$-variation.
To see this, let us pick a function $g:S^{3}\rightarrow\mathbb{R}$
and consider:

\[
\delta(\int\sqrt{g}d^{3}x\epsilon^{\dagger}\gamma^{\mu}(\partial_{\mu}g)\lambda)=\int\sqrt{g}d^{3}x\epsilon^{\dagger}\gamma^{\mu}(\partial_{\mu}g)(i\gamma^{\nu}(-\frac{1}{2}\epsilon_{\nu\rho\sigma}F^{\rho\sigma}+\partial_{\nu}\sigma)-(D+\sigma))\epsilon)
\]

\[
=\int\sqrt{g}d^{3}x(\partial_{\mu}g)\bigg(i(g^{\mu\nu}+i\epsilon^{\mu\nu\tau}v_{\tau})(-\frac{1}{2}\epsilon_{\nu\rho\sigma}F^{\rho\sigma}+\partial_{\nu}\sigma)-v_{\mu}(D+\sigma)\bigg)
\]
The term involving $D+\sigma$ is proportional to $v^{\mu}\partial_{\mu}g$
and vanishes if we impose that $g$, like $f$, is constant along
the fibers. The remaining terms can be expanded to give:

\[
\int\sqrt{g}d^{3}x(\partial_{\mu}g)\bigg(-\frac{i}{2}\epsilon^{\mu\rho\sigma}F_{\rho\sigma}+F^{\mu\nu}v_{\nu}+i\partial^{\mu}\sigma-i\epsilon^{\mu\nu\rho}\partial_{\nu}\sigma v_{\rho}\bigg)
\]
Integrating by parts, the first and last terms can be seen to vanish,
and the others give

\[
\int\sqrt{g}d^{3}x\bigg(A^{\mu}(\frac{1}{2}\nabla^{\nu}(v_{\mu}\partial_{\nu}g-v_{\nu}\partial_{\mu}g))-i\sigma(\nabla^{2}g)\bigg)
\]
The quantity multiplying $A^{\mu}$ can be expanded as:

\[
\nabla^{\nu}v_{\mu}(\partial_{\nu}g)+v_{\mu}\nabla^{2}g-(\nabla^{\nu}v_{\nu})\partial_{\mu}g-v_{\nu}\nabla^{\nu}\partial_{\mu}g
\]
Using $\nabla_{\mu}v_{\nu}=\epsilon_{\mu\nu\rho}v^{\rho}$, this can
be simplified to:

\[
v_{\mu}\nabla^{2}g-\nabla_{\mu}(v_{\nu}\nabla^{\nu}g)
\]
The second term vanishes when we impose that $g$ is constant along
the fibers, and we are left with:

\[
\int\sqrt{g}d^{3}x(\nabla^{2}g)(v^{\mu}A_{\mu}-i\sigma)
\]
which agrees with the exponent of the operator $\mathcal{O}_{f_{o}}$
above, provided we can find a $g$ such that:

\[
f_{o}=\nabla^{2}g.
\]
This clearly requires $\int_{S^{3}}\sqrt{g}d^{3}xf_{o}=0$, so that
one cannot use this trick to remove the constant part of $f$. However,
if this condition is met, then the equation can be solved, and $g$
will indeed be constant along the fibers.%
\footnote{This can be seen most easily by working with a mode expansion $f_{o}=\sum_{\ell,m,n}c_{\ell,m,n}Y_{\ell,m,n}$,
where $Y_{\ell,m,n}$ are spherical harmonics on $S^{3}$, satisfying
$\nabla^{2}Y_{\ell,m,n}=-\ell(\ell+2)Y_{\ell,m,n}$ and $v^{\mu}\partial_{\mu}Y_{\ell,m,n}=imY_{\ell,m,n}$.
Then $g=\sum_{\ell,m,n}\frac{1}{\ell(\ell+2)}c_{\ell,m,n}Y_{\ell,m,n}$,
which is well defined since $f$ has no $\ell=0$ component by assumption,
and, like $f$, has $c_{\ell,m,n}=0$ for all $m\neq0$.%
} This proves that the non-constant part of the operator can be discarded,
as $\delta$-exact terms do not affect the path integral, i.e. we
can replace a Wilson loop localized on a loop by one that is smeared
uniformly over the entire $S^{3}$; these differ only by a $\delta$-exact
insertion.

Thus without a loss of generality we can restrict to the case $f=\frac{q}{\pi}$
(a constant). Then the background vector multiplet we must couple
the flavor symmetry current to is given by:

\[
F=2q\star v,\;\;\; D=-2iq
\]
Since $dv=2\star v$, the corresponding gauge field can be taken to
be

\[
A=qv
\]
The $\delta$-exact gauged action of a chiral multiplet of conformal
dimension $1/2$ is given by \cite{Kapustin:2009kz}

\[
S_{\delta}=\int\sqrt{g}d^{3}x\bigg(\phi^{\dagger}(-D_{\mu}D^{\mu}+\sigma^{2}+iD+\frac{3}{4})\phi+\psi^{\dagger}(i\gamma^{\mu}D_{\mu}-i\sigma)\psi+F^{\dagger}F\bigg)
\]
Let us couple this to an ordinary BPS background, with $\sigma=-D=\sigma_{o}$,
as well as the background vector multiplet described above, with $A=qv$
and $D=-2iq$. We find:

\[
S_{\delta}=\int\sqrt{g}d^{3}x\bigg(\phi^{\dagger}(-\nabla^{2}-2iqv^{\mu}\partial_{\mu}+q^{2}+{\sigma_{o}}^{2}-i\sigma_{o}+2q+\frac{3}{4})\phi+\psi^{\dagger}(i\gamma^{\mu}\nabla_{\mu}-q\gamma^{\mu}v_{\mu}-i\sigma_{o})\psi+F^{\dagger}F\bigg)
\]

The bosonic operator has the form:

\[
\mathcal{O}_{B}=-\nabla^{2}+aiv^{\mu}\partial_{\mu}+b
\]
with $a=-2q$ and $b=q^{2}+{\sigma_{o}}^{2}-i\sigma_{o}+2q+\frac{3}{4}$,
which has determinant \cite{Kapustin:2009kz}

\[
\sqrt{\det\mathcal{O}_{B}}=\prod_{\ell=0}^{\infty}\bigg(\prod_{m=-\ell/2}^{\ell/2}\bigg(\ell(\ell+2)-2am+b\bigg)\bigg)^{\ell+1}
\]

\[
=\prod_{\ell=0}^{\infty}\bigg(\prod_{m=-\ell/2}^{\ell/2}\bigg(\ell(\ell+2)+4qm+q^{2}+{\sigma_{o}}^{2}-i\sigma_{o}+2q+\frac{3}{4}\bigg)\bigg)^{\ell+1}
\]
Meanwhile, for the fermions, the operator has the form:

\[
\mathcal{O}_{F}=i\gamma^{\mu}\nabla_{\mu}+ic\gamma^{\mu}v_{\mu}+d
\]
with $c=-q$ and $d=-i\sigma_{o}$, which has determinant:

\[
\det\mathcal{O}_{F}=\prod_{\ell=0}^{\infty}\bigg((\ell-c-d+3/2)(\ell+c-d+3/2)\prod_{m=-\ell/2}^{\ell/2-1}\bigg(\ell(\ell+2)-4cm-2c+d+c^{2}-d^{2}+3/4\bigg)\bigg)^{\ell+1}
\]

\[
=\prod_{\ell=0}^{\infty}\bigg((\ell+q+i\sigma_{o}+3/2)(\ell-q+i\sigma_{o}+3/2)\prod_{m=-\ell/2}^{\ell/2-1}\bigg(\ell(\ell+2)+4qm+2q-i\sigma_{o}+q^{2}+{\sigma_{o}}^{2}+3/4\bigg)\bigg)^{\ell+1}
\]
We see that most modes cancel, and we are left with:

\[
Z_{1-loop}=\frac{\det\mathcal{O}_{F}}{\sqrt{\det\mathcal{O}_{B}}}=\prod_{\ell=0}^{\infty}\bigg(\frac{(\ell+q+i\sigma_{o}+3/2)(\ell-q+i\sigma_{o}+3/2)}{(\ell(\ell+2)-2q\ell+2q-i\sigma_{o}+q^{2}+{\sigma_{o}}^{2}+3/4)}\bigg)^{\ell+1}
\]

\[
=\prod_{\ell=0}^{\infty}\bigg(\frac{(\ell+q+i\sigma_{o}+3/2)(\ell-q+i\sigma_{o}+3/2)}{(\ell+1+q+i\sigma_{o}+\frac{1}{2})(\ell+1+q-i\sigma_{o}-\frac{1}{2})}\bigg)^{\ell+1}
\]

\[
=\prod_{\ell=0}^{\infty}\bigg(\frac{\ell+i(\sigma_{o}+iq)+3/2}{\ell-i(\sigma_{o}+iq)+1/2}\bigg)^{\ell+1}
\]
Note that $\sigma_{o}$ and $\alpha$ appear in the combination $\sigma+iq$,
so that we can just make this replacement in the ordinary one-loop
determinant to find: 
\[
Z_{1-loop}=e^{\ell(\frac{1}{2}+i\sigma-q)}
\]
In fact, this computation goes through with minimal changes for chiral
multiplets of arbitrary dimension (one merely shifts $\sigma_{o}$
by an imaginary amount), and we find:

\[
Z_{1-loop}=e^{\ell(1-\Delta+i\sigma-q)}
\]
Note we have obtained the same result as in the indirect argument
above. We will now turn to an even more explicit argument, where we
do not smear out the defect over the sphere but instead work directly
with a (regularized) delta function background.

\subsection{\label{sub:explicit_computation_in_the_delta_function_background}Method
$3$: explicit computation in a singular background}

Let us now focus on the specific case where the function $\tf$ is
a delta function, corresponding to the dual of an ordinary (unsmeared)
Wilson loop. Although we have argued that one can replace the delta
function by a constant which has the same integral over $S^{2}$,
we would like to gain better physical insight into the vortex loop
by explicitly finding the modes in a singular background. For simplicity
we will focus on the case where the matter has canonical dimension,
although it is straightforward to generalize this.

Recall that the smeared vortex loop on $S^{3}$ can be obtained by
coupling to a background $F=2\pi f\star v$ and $D=-2\pi if$, where
$f$ is some function on $S^{3}$ constant along the fibers of the
Hopf fibration. Here we compute the modes and $1$-loop determinant
explicitly in the case where $f$ is a (infinitessimally smeared)
delta function supported on a single fiber.

We will work in toroidal coordinates on $S^{3}$, with $\eta\in[0,\pi/2]$
and $\theta$ and $\phi$ in $[0,2\pi)$. Explicitly, we can relate
these coordinates to the unit sphere $S^{3}\subset\mathbb{R}^{4}$
via:

\[
x=\cos\eta\cos\theta,\;\;\; y=\cos\eta\sin\theta,\;\;\; z=\sin\eta\cos\phi,\;\;\; w=\sin\eta\sin\phi
\]
The surfaces of constant $\eta$ are torii, which degenerate to great
circles at $\eta=0,\frac{\pi}{2}$.

The usual round metric takes the following form in these coordinates:

\[
ds^{2}=d\eta^{2}+\sin^{2}\eta d\theta^{2}+\cos^{2}\eta d\phi^{2}
\]

\[
\Rightarrow\nabla^{2}=\frac{1}{\sin\eta\cos\eta}\frac{\partial}{\partial\eta}\sin\eta\cos\eta\frac{\partial}{\partial\eta}+\frac{1}{\sin^{2}\eta}\frac{\partial^{2}}{\partial\theta^{2}}+\frac{1}{\cos^{2}\eta}\frac{\partial^{2}}{\partial\phi^{2}}
\]

The Killing vector $v$ is given in these coordinates by:

\[
v=\frac{\partial}{\partial\theta}+\frac{\partial}{\partial\phi}
\]
or, as a $1$-form, by:

\[
\tilde{v}=\sin^{2}\eta d\theta+\cos^{2}\eta d\phi
\]
It satisfies:

\[
dv=2\cos\eta\sin\eta d\eta\wedge d\theta-2\cos\eta\sin\eta d\eta\wedge d\phi=2\star v
\]
We take the defect to be supported on the great circle at $\eta=0$.
We define a regularized delta function supported on the loop by:

\[
f=\frac{g(\eta/\epsilon)}{2\pi\epsilon\sin\eta\cos\eta}
\]
where $g(x)$ has support in $0\leq x\lesssim1$ and integrates to
$1$ on $[0,\frac{\pi}{2\epsilon}]$, so that $f$ integrates to $2\pi$.
This approaches a delta function supported on the great circle at
$\eta=0$ for $\epsilon\rightarrow0$. Then the background field-strength
we need to consider is 
\[
F=2\pi q\; f\star v=\frac{q}{\epsilon}g(\eta/\epsilon)d\eta\wedge(d\theta-d\phi)
\]
which is solved by a background vector potential 
\[
A=qG(\eta/\epsilon)d\theta-q(G(\eta/\epsilon)-1)d\phi
\]
where $G'=g$ with $G(0)=0$, so that $G(x)\rightarrow1$ for large
$x$, and we pick the constants so that this is everywhere well-defined.
We also have a background auxiliary scalar 
\[
D=-2\pi iqf=-\frac{iq\; g(\eta/\epsilon)}{\epsilon\sin\eta\cos\eta}
\]

\subsubsection{Bosons}

First consider the bosonic operator on $S^{3}$:

\[
\mathcal{O}_{B}=-D_{\mu}D^{\mu}+\sigma^{2}+iD+\frac{3}{4}
\]
We will couple to a defect background, as above, as well as an ordinary
$\sigma=-D=\sigma_{o}$ background. Using the form of $D$ above,
along with:

\[
A^{\mu}\partial_{\mu}=\bigg(\frac{q\; G(\eta/\epsilon)}{\sin^{2}\eta}\frac{\partial}{\partial\theta}-\frac{q(G(\eta/\epsilon)-1)}{\cos^{2}\eta}\frac{\partial}{\partial\phi}\bigg)
\]

\[
A^{2}=\bigg(\frac{q^{2}G(\eta/\epsilon)^{2}}{\sin^{2}\eta}+\frac{q^{2}(G(\eta/\epsilon)-1)^{2}}{\cos^{2}\eta}\bigg)
\]
and looking for an eigenfunction $\mathcal{O}_{B}\phi=\lambda\phi$
of the form $\phi=f(\eta)e^{im\theta+in\phi}$, we get the following
equation:

\[
-\frac{1}{\sin\eta\cos\eta}\frac{d}{d\eta}\bigg(\sin\eta\cos\eta\frac{df}{d\eta}\bigg)+\bigg(\frac{(m+qG)^{2}}{\sin^{2}\eta}+\frac{(n-q(G-1))^{2}}{\cos^{2}\eta}+\frac{qg(\eta/\epsilon)}{\epsilon\sin\eta\cos\eta}+\frac{3}{4}+{\sigma_{o}}^{2}-i\sigma_{o}-\lambda\bigg)f=0
\]
Before solving this equation, let us consider the fermions, as we
will see their components satisfy a very similar equation.

\subsubsection{Fermions}

The operator in this case is

\[
\mathcal{O}_{F}=iD\!\!\!\!/\;-i\sigma
\]
It is convenient to use a left-invariant vielbein. One computes this
in toroidal coordinates as:

\[
e_{i}^{L}=\left\{ \begin{array}{cc}
{\displaystyle \sin(\theta+\phi)\frac{\partial}{\partial\eta}+\cos(\theta+\phi)(\cot\eta\frac{\partial}{\partial\theta}-\tan\eta\frac{\partial}{\partial\phi})} & i=1\\
\\
{\displaystyle -\cos(\theta+\phi)\frac{\partial}{\partial\eta}+\sin(\theta+\phi)(\cot\eta\frac{\partial}{\partial\theta}-\tan\eta\frac{\partial}{\partial\phi})} & i=2\\
\\
{\displaystyle \frac{\partial}{\partial\theta}+\frac{\partial}{\partial\phi}} & i=3
\end{array}\right.
\]
Then the Dirac operator can be written as:

\[
i\dirac=\left(\begin{array}{cc}
i\partial_{3}-\frac{3}{2} & i\partial_{1}+\partial_{2}\\
i\partial_{1}-\partial_{2} & -i\partial_{3}-\frac{3}{2}
\end{array}\right)
\]
which, in toroidal coordinates, becomes:

\[
i\dirac=\left(\begin{array}{cc}
{\displaystyle i\frac{\partial}{\partial\theta}+i\frac{\partial}{\partial\phi}-\frac{3}{2}} & {\displaystyle e^{-i(\theta+\phi)}\bigg(-\frac{\partial}{\partial\eta}+i\bigg(\cot\eta\frac{\partial}{\partial\theta}-\tan\eta\frac{\partial}{\partial\phi}\bigg)\bigg)}\\
{\displaystyle e^{i(\theta+\phi)}\bigg(\frac{\partial}{\partial\eta}+i\bigg(\cot\eta\frac{\partial}{\partial\theta}-\tan\eta\frac{\partial}{\partial\phi}\bigg)\bigg)} & {\displaystyle -i\frac{\partial}{\partial\theta}-i\frac{\partial}{\partial\phi}-\frac{3}{2}}
\end{array}\right)
\]
We should also couple to the gauge field, which amounts to the replacement
$\frac{\partial}{\partial\theta}\rightarrow\frac{\partial}{\partial\theta}+iqG,\frac{\partial}{\partial\phi}\rightarrow\frac{\partial}{\partial\phi}-iq(G-1)$,
as well as the constant $\sigma_{o}$. Then if we look for eigenspinors
of the form

\[
\psi=e^{im\theta+in\phi}\left(\begin{array}{c}
\psi_{1}(\eta)\\
e^{i(\theta+\phi)}\psi_{2}(\eta)
\end{array}\right)
\]
we get the following coupled first-order equations for $\psi_{1}$
and $\psi_{2}$:

\begin{align}
(\frac{d}{d\eta}-(m+qG)\cot\eta+(n-q(G-1))\tan\eta)\psi_{1} & =-(m+n+q-\lambda-i\sigma_{o}+\frac{1}{2})\psi_{2}\label{ce}\\
\notag(-\frac{d}{d\eta}-(m+qG+1)\cot\eta+(n-q(G-1)+1)\tan\eta)\psi_{2} & =(m+n+q+\lambda+i\sigma_{o}+\frac{3}{2})\psi_{1}
\end{align}
Solving for $\psi_{2}$ using the first equation and plugging into
the second one, we get a second order equation in terms of $\psi_{1}$
alone:

\[
\bigg(-\frac{d}{d\eta}-(m+qG+1)\cot\eta+(n-q(G-1)+1)\tan\eta\bigg)\bigg(\frac{d}{d\eta}-(m+qG)\cot\eta=(n-q(G-1))\tan\eta\bigg)\psi_{1}=
\]

\begin{equation}
=-(m+n+q-\lambda-i\sigma_{o}+\frac{1}{2})(m+n+q+\lambda+i\sigma_{o}+\frac{3}{2})\psi_{1}.\label{psie}
\end{equation}
This can be rearranged to 
\[
-\frac{1}{\sin\eta\cos\eta}\frac{d}{d\eta}\bigg(\sin\eta\cos\eta\frac{d\psi_{1}}{d\eta}\bigg)+\bigg(\frac{(m+qG)^{2}}{\sin^{2}\eta}+\frac{(n-q(G-1))^{2}}{\cos^{2}\eta}+\frac{qg(\eta/\epsilon)}{\epsilon\sin\eta\cos\eta}-(\lambda+i\sigma_{o}+\frac{1}{2})^{2}+1\bigg)\psi_{1}=0
\]
which is precisely the same equation as satisfied by the bosonic modes,
with the following relation between the eigenvalues:

\begin{equation}
\lambda_{B}+(i\sigma_{o}+\frac{1}{2})^{2}=(\lambda_{F}+i\sigma_{o}+\frac{1}{2})^{2}\label{eigs}
\end{equation}

This is a quadratic equation for $\lambda_{F}$ with two solutions
${\lambda_{F}}_{\pm}$. Their product can be read off as the constant
term, and one can see that this is simply $\lambda_{B}$.

In principle, we can take $\psi_{1}$ to be the scalar eigenfunction
for a given choice of $\lambda_{B}$, take one of the solutions ${\lambda_{F}}_{\pm}$
to the equation above, and plug these into the first equation in (\ref{ce})
to solve for $\psi_{2}$. For each $\lambda_{B}$ and pair of solutions
${\lambda_{F}}_{\pm}$ for which this procedure goes through, we can
see that the contribution to the partition function, ${\lambda_{F}}_{+}{\lambda_{F}}_{-}/\lambda_{B}=1$,
is trivial.

However, there are two exceptions we must be more careful with. First,
we must also allow solutions with $\psi_{1}=0$, but $\psi_{2}$ non-vanishing.
Then we see that (\ref{ce}) reduces to:

\begin{align*}
(m+n+q-\lambda-i\sigma_{o}+\frac{1}{2})\psi_{2} & =0\\
(-\frac{d}{d\eta}-(m+qG+1)\cot\eta+(n-q(G-1)+1)\tan\eta)\psi_{2} & =0
\end{align*}
Thus the eigenvalue in these cases is $\lambda=m+n+q-i\sigma_{o}+\frac{1}{2}$,
and $\psi_{2}$ satisfies:

\[
\frac{d}{d\eta}\log\psi_{2}=-(m+qG+1)\cot\eta+(n-q(G-1)+1)\tan\eta=\left\{ \begin{array}{cc}
-(m+1)\cot\eta+... & \mbox{near \ensuremath{\eta=0}}\\
-(m+q+1)\cot\eta+(n+1)\tan\eta & \mbox{ in the bulk}
\end{array}\right.
\]

\[
\Rightarrow\psi_{2}=\left\{ \begin{array}{cc}
\sin^{-(m+1)}\eta+... & \mbox{near \ensuremath{\eta=0}}\\
\sin^{-(m+q+1)}\eta\cos^{-(n+1)}\eta & \mbox{ in the bulk}
\end{array}\right.
\]
Regularity at the endpoints implies that $m$ and $n$ should be negative
integers. But then for $-(q+1)<m<0$, the bulk solution is singular
as it approaches the loop. We will return to this point in a moment.
These solutions correspond to extra fermionic modes that we have not
accounted for before, so their eigenvalues should be included in the
numerator of the partition function.

The other exception occurs when the differential operator acting on
$\psi_{1}$ in the first equation in (\ref{ce}) annihilates our choice
of $\psi_{1}$. Then we must pick $\psi_{2}=0$, and there will not
be two choices of $\lambda_{F}$, but only one, and so the cancellation
with the corresponding bosonic mode will not be complete. We see that
in this case (\ref{ce}) gives:

\begin{align*}
(\frac{d}{d\eta}-(m+qG)\cot\eta+(n-q(G-1))\tan\eta)\psi_{1} & =0\\
\notag(m+n+q+\lambda+i\sigma_{o}+\frac{3}{2})\psi_{1} & =0
\end{align*}
Now we find $\lambda=-(m+n+q+i\sigma_{o}+\frac{3}{2})$, and $\psi_{1}$
satisfies:

\[
\frac{d}{d\eta}\log\psi_{1}=(m+qG)\cot\eta-(n-q(G-1))\tan\eta=\left\{ \begin{array}{cc}
m\cot\eta+... & \mbox{near \ensuremath{\eta=0}}\\
(m+q)\cot\eta-n\tan\eta & \mbox{ in the bulk}
\end{array}\right.
\]

\[
\Rightarrow\psi_{1}=\left\{ \begin{array}{cc}
\sin^{m}\eta+... & \mbox{near \ensuremath{\eta=0}}\\
\sin^{(m+q)}\eta\cos^{n}\eta & \mbox{ in the bulk}
\end{array}\right.
\]
Now regularity at the endpoints forces $m$ and $n$ to be nonnegative
integers, and the bulk solutions are singular for $0\leq m<-q$. These
are modes for which there is only one solution, say ${\lambda_{F}}_{+}$,
to the equation (\ref{eigs}), and so the cancellation with the bosons
is not complete. The net contribution is ${\lambda_{F}}_{+}/\lambda_{B}=1/{\lambda_{F}}_{-}$,
and one can read this off from (\ref{psie}) as ${\lambda_{F}}_{-}=m+n+q-i\sigma_{o}+\frac{1}{2}$.

Putting this together, we see all modes cancel out of the partition
function except for the special cases noted above, and these give:

\[
Z=\frac{\prod_{m,n<0}(m+n+q-i\sigma_{o}-\frac{1}{2})}{\prod_{m,n\geq0}(m+n+q-i\sigma_{o}+\frac{1}{2})}
\]

\[
=\prod_{\ell=0}^{\infty}\bigg(\frac{-\ell+q-i\sigma_{o}-\frac{3}{2}}{\ell+q-i\sigma_{o}+\frac{1}{2}}\bigg)^{\ell+1}
\]

\[
=e^{\ell(\frac{1}{2}+q-i\sigma_{o})}
\]

Thus we have obtained the result for the $1$-loop determinant without
ever having to solve the second order differential equation. However,
note that this function may have poles, e.g. when $q=\frac{1}{2}$.
To get a better understanding of the origin of these singularities,
we will now solve the differential equation explicitly.

\subsubsection{Solving the eigenvalue equation}

Consider the following second order equation, which has come up for
both the bosons and fermions:

\[
-\frac{1}{\sin\eta\cos\eta}\frac{d}{d\eta}\bigg(\sin\eta\cos\eta\frac{df}{d\eta}\bigg)+\bigg(\frac{(m+qG)^{2}}{\sin^{2}\eta}+\frac{(n-q(G-1))^{2}}{\cos^{2}\eta}+\frac{qg(\eta/\epsilon)}{\epsilon\sin\eta\cos\eta}+1-(\ell+1)^{2}\bigg)f=0
\]
where, for later convenience, we have written the eigenvalue in terms
of a parameter $\ell$. This is related to the eigenvalues by:

\[
\lambda_{B}=(\ell+1)^{2}-(i\sigma_{o}+\frac{1}{2})^{2},\;\;\;\;\;\lambda_{F}{}_{\pm}=\pm(\ell+1)-i\sigma_{o}-\frac{1}{2}
\]

\paragraph{Near Loop Region}

Let us start by focusing on the region $0<\eta\lesssim\epsilon$,
as this will determine what boundary conditions to impose on the bulk
solution. We start by defining $\xi=\eta/\epsilon$, and expanding
the equation above to leading order in $\epsilon$:

\[
-\frac{1}{\xi}\frac{d}{d\xi}\bigg(\xi\frac{df}{d\xi}\bigg)+\bigg(\frac{(m+qG(\xi))^{2}}{\xi^{2}}+\frac{qg(\xi)}{\xi}\bigg)f=0
\]
One solution can be immediately obtained, independent of the functional
form of $g$ and $G$, by noting that this equation follows from the
first order equation:

\[
\frac{df}{d\xi}=\frac{m+qG}{\xi}f
\]
as can be easily checked using $G'=g$. Then, using $G\rightarrow0$
as $\chi\rightarrow0$ and $G\rightarrow1$ as $\chi\rightarrow\infty$,
we see that:

\[
f(\xi)\sim\left\{ \begin{array}{cc}
\xi^{m} & \xi\rightarrow0\\
\xi^{(m+q)} & \xi\rightarrow\infty
\end{array}\right.
\]
For $m\geq0$, this is the regular solution we want. We can match
with the solution in the bulk by looking at the behavior in the the
large $\xi$ region, and we see that we should take the bulk solution
which goes as $f(\eta)\sim\eta^{(m+q)}$. Note that if $q<0$, for
$0\leq m<-q$, the bulk solution would appear to be singular right
up until we reach the near loop region, at which point the presence
of the defect modifies the solution to go as $\eta^{-m}$ and be regular.
For $m\geq-q$, we take the regular solution, as in the absence of
a defect. Note that this behavior does not depend on the precise functional
form of $g$, only that it correctly reproduces a delta function in
the $\epsilon\rightarrow0$ limit.

It remains to understand what happens when $m<0$. Here, the first
order equation does not have any non-trivial solutions regular at
$\xi=0$, so we must return to the second order equation. Then we
can find a regular solution, but it appears it depends non-trivially
on $g$ and $G$. Specifically, we find:

\[
f(\xi)\sim\left\{ \begin{array}{cc}
\xi^{m} & \xi\rightarrow0\\
A(g,m,q)\xi^{|m+q|}+B(g,m,q)\xi^{-|m+q|} & \xi\rightarrow\infty
\end{array}\right.
\]
for some constants which depend non-trivially on $g$, $m$, and $q$,
and in particular are generically both non-zero, unlike in the previous
case.

Nevertheless, if we reinstate the $\epsilon$ dependence, we see that
to match with the bulk, we should take the solution there to go as:

\[
f(\eta)\sim A(g,m,q)\eta^{|m+q|}\epsilon^{-|m+q|}+B(g,m,q)\eta^{-|m+q|}\epsilon^{|m+q|}
\]
and for $\epsilon\rightarrow0$, the first term will dominate, and
so we should take the regular solution.

Thus we should always take the regular solution, $f\sim\eta^{|m+q|}$,
except in the case where the coefficient of the regular part is precisely
zero, which happens only when $0\leq m<-q$, and in these cases we
take the singular solution $f\sim\eta^{-|m+q|}$.

\paragraph{Bulk}

Since $G=1$ and $g=0$ everywhere except an infinitessimal region
near $\eta=0$, in the bulk the equation reduces to:

\[
-\frac{1}{\sin\eta\cos\eta}\frac{d}{d\eta}\bigg(\sin\eta\cos\eta\frac{df}{d\eta}\bigg)+\bigg(\frac{(m+q)^{2}}{\sin^{2}\eta}+\frac{n^{2}}{\cos^{2}\eta}+1-(\ell+1)^{2}\bigg)f=0
\]
It is convenient to look for an $f$ of the form:

\[
f(\eta)=\sin^{\tm}\eta\cos^{\tn}\eta h(\sin^{2}\eta)
\]
for $\tm,\tn$ we will choose in a moment. Plugging this in, we find
the following equation for $h(x)$:

\[
x(1-x)h''+(\tm+1-(\tm+\tn+2)x)h'-\frac{1}{4}\bigg(\frac{(m+q)^{2}-\tm^{2}}{x}+\frac{(n-q)^{2}-\tn^{2}}{1-x}+(\tm+\tn+1)^{2}-(\ell+1)^{2}\bigg)h=0
\]
The greatest simplification is achieved by setting $\tm^{2}=(m+q)^{2}$
and $\tn^{2}$, with the sign of $\tm$ and $\tn$ to be fixed later.
Then we are left with the hypergeometric equation:

\[
x(1-x)h''+(c-(a+b+1)x)h'-abh=0
\]
where:

\[
c=\tm+1,\;\;\; a+b=\tm+\tn+1,\;\;\; ab=\frac{1}{4}((\tm+\tn+1)^{2}-(\ell+1)^{2})
\]

\[
\Rightarrow a,b=\frac{1}{2}(\tm+\tn+1\pm(\ell+1))
\]
The solutions can be written in terms of hypergeometric functions
$\,_{2}F_{1}(a,b;c;x)$:

\[
\,_{2}F_{1}(a,b;c;x)=\sum_{n=0}^{\infty}\frac{(a)_{n}(b)_{n}}{(c)_{n}n!}x^{n}
\]
where $(a)_{n}=a(a+1)...(a+n-1)$. Provided this is well-defined%
\footnote{Specifically, if $c$ is a negative integer, we must have $a$ or
$b$ to be a negative integer greater than or equal to $c$ for this
series to make sense.%
}, it converges for all $|x|<1$, and for $|x|=1$ if $\mbox{Re}(c-a-b)>0$.

In general, there are two linearly independent solutions, but if we
impose regularity at the boundary at $x=1$ this restricts us to the
solution:

\[
h(x)=\,_{2}F_{1}(a,b;a+b+1-c;1-x)
\]

\[
=\frac{\Gamma(a+b+1-c)\Gamma(1-c)}{\Gamma(a+1-c)\Gamma(b+1-c)}\,_{2}F_{1}(a,b;c;x)+\frac{\Gamma(a+b+1-c)\Gamma(c-1)}{\Gamma(a)\Gamma(b)}x^{1-c}\,_{2}F_{1}(a+1-c,b+1-c;2-c;x)
\]

\[
=C_{1}(a,b,c)\,_{2}F_{1}(a,b;c;x)+C_{2}(a,b,c)\,_{2}F_{1}(a+1-c,b+1-c;2-c;x)
\]
we should also set $\tn=|n|$ to ensure $f(\eta)$ is regular.

Now we need to fix the behavior at the other endpoint. We have:

\[
f(\eta)\sim C_{1}(a,b,c)\eta^{\tm}+C_{2}(a,b,c)\eta^{-\tm}
\]
Thus, depending on the behavior we want here, we should fix the relative
values of the $C_{i}$ which will impose a condition on $\ell$ and
restrict us to a discrete spectrum. Actually, since we are free to
choose the sign of $\tm$, we can simply pick $\tm$ so that the behavior
we want is $\eta^{\tm}$, and then the condition is always $C_{2}=0$.
This is ensured if the argument of one of the Gamma functions in the
denominator in $C_{2}$ is set to a negative integer, which determines:

\[
a,b=\frac{1}{2}(\tm+|n|+1\pm(\ell+1))=-k,\;\;\; k\in\mathbb{Z}_{>0}
\]

\[
\Rightarrow\lambda_{B}=(\ell+1)^{2}+(\sigma_{o}-\frac{i}{2})^{2}=(\tm+|n|+2k+1)^{2}+(\sigma_{o}-\frac{i}{2})^{2}
\]

\[
\lambda_{F}=\pm(\ell+1)-i\sigma_{o}-\frac{1}{2}=\pm(\tm+|n|+2k+1)-i\sigma_{o}-\frac{1}{2}
\]

Finally, we recall that the correct choice of $\tm$ was found above
to be:

\[
\tm=\left\{ \begin{array}{cc}
-|m-q| & 0\leq m<-q\\
|m-q| & \mbox{ else}
\end{array}\right.
\]
Then the set of eigenvalues is given by taking the expressions above
for all integers $m,n$ and non-negative integers $k$.

Actually, as above, we need to be more careful with the fermions.
Specifically, there will be eigenvalues in addition to these, as well
as some that we have to throw out, corresponding to the cases where,
respectively, the top and bottom components of the fermion vanish.
These are precisely the exceptions noted in the previous section,
from which the entire contribution to the partition function comes.

One interesting property of these eigenvalues is that, since $\tm$
may be negative, the eigenvalues may become zero, and the bosonic
eigenvalues may even be negative. The first place these zero modes
can occur is for $q=-\frac{1}{2}$, in which case for $m=k=\sigma_{o}=0$,
$\lambda_{B}$ and $\lambda_{F}^{+}$ are both zero. Actually this
fermionic eigenvalue is one of the spurious ones that we should throw
out, and so in fact there is a single bosonic zero mode which results
in a pole in the $1$-loop partition function. For larger $q$, one
finds negative bosonic eigenvalues, which become difficult to make
sense of in the path integral. We will take the viewpoint that the vortex loop 
operator is only properly defined for $q>-\frac{1}{2}$ (and, when
the matter content is in a self-conjugate representation of the relevant
flavor symmetry, this also forces $q<\frac{1}{2}$), although the
naive result for the determinant gives a natural analytic continuation
of this quantity which may have some physical relevance.

\section{\label{sec:Duality-with-defect-operators}Duality with Vortex Loop
Operators}

Let us now turn to some applications of the vortex loop operator. Since
this operator is defined by an explicit procedure applied to a global
$U(1)$ symmetry, then if we know what this symmetry maps to across
a duality, we obtain an identification of loop operators on each
side of the duality. This provides a new operator mapping across duality,
although it does not provide any essentially new information beyond
the mapping of global symmetries.

One inter?sting case is when a $U(1)_{J}$ symmetry on one side of
the duality is identified with a flavor symmetry on the dual. This
occurs, for example, in $3D$ mirror symmetry. Let us take the simplest
case, the duality between $\mathcal{N}=4$ SQED with one flavor and
a free hypermultiplet. Note that there are Wilson loops on the SQED
side, but it is less clear what the corresponding loop operators on
the dual side are, since there is no gauge group.

However, the results above provide the answer: the Wilson loop on
the SQED side are the same as defects in the global $U(1)_{J}$ symmetry.
Mirror symmetry dictates that this symmetry is identified with a $U(1)_{V}$
flavor symmetry under which the two chirals in the hypermultiplet
have charge $\pm1$. Then the Wilson loop in SQED simply maps to a
defect operator in this flavor symmetry, supported on the same loop.

At the level of the matrix model, this follows from the identity:

\[
\int d\lambda\frac{e^{2\pi i\eta\lambda}e^{2\pi q\lambda}}{2\cosh(\pi\lambda)}=\frac{1}{2\cosh\pi(\eta-iq)}
\]
which simply follows from extending the usual self-Fourier-transform
property of $1/\cosh$ to the entire complex plane.

It is worth noting that the integral on the LHS only converges for
$|q|<\frac{1}{2}$. As remarked above, this is precisely the range
in which the defect operator is also well-defined. The divergence
on the LHS as $|q|\rightarrow\frac{1}{2}$ is reflected on the RHS
as a bosonic zero mode developing in the defect background. The RHS
gives a natural analytic continuation of this quantity to $|q|>\frac{1}{2}$,
but it is not clear what, if any, physical relevance this has.

We can also consider mirror symmetry applied to SQED with $N_{f}$
flavors. Here the dual is a quiver theory with gauge group $U(1)^{N_{f}}/U(1)_{diag}\cong U(1)^{N_{f}-1}$,
with $N_{f}$ bifundamental flavors $(q_{a}\tilde{q}_{a})$ charged
as $(1,-1,0,...),(0,1,-1,...),...,(-1,0,...,1)$, $N_{f}$ neutral
chirals $S_{a}$, and a superpotential $\sum_{a}q_{a}S_{a}\tilde{q}_{a}$.
Here the $U(1)_{J}$ symmetry of $SQED$ maps to a flavor symmetry
under which the all the $q_{a}$ have charge $\frac{1}{N_{f}}$. In
particular, the Wilson loop in SQED maps to a defect in this symmetry.
Note that, because of the extra factor of $\frac{1}{N_{f}}$ in this
mapping, the restriction on the defect charge now allows us to consider
Wilson loops of charge up to $\pm\frac{N_{f}}{2}$. This coincides
with the fact that, in the matrix model, the Wilson loop expectation
value now converges for this wider range of charges because of the
increased damping in the integrand from the factor $(2\cosh(\pi\lambda))^{-N_{f}}$.

Finally, we note that we can also apply this type of argument to theories
with $U(N)$ gauge symmetries for $N>1$, provided we restrict only
to Wilson loops in the overall $U(1)$ of the gauge group. For example,
in the $U(N)$ versions of Aharony (\cite{Aharony:1997gp}) and Giveon-Kutasov
dualities (\cite{Giveon:2008zn}), it is known that the $U(1)_{J}$
current maps to itself, up to a flip of sign, and so a Wilson loop
in the overall $U(1)$ of the gauge group must map to the same Wilson
loop, with the opposite charge. One can also consider such abelian
Wilson loops in non-abelian mirror symmetry, where they will map to
flavor defects as above.

\section{Discussion}

We have defined a set of abelian vortex loop operators which exist
in any abelian gauge theory in $2+1$ dimensions. The definition can
be extended to conformal field theories with abelian global symmetries
by weakly gauging the symmetry currents. Witten's $SL(2,\mathbb{Z})$
action for this class of theories naturally extends to loop operators.
In fact, abelian vortex loops are the S-duals of the ordinary Wilson
loops. Alternatively, a Wilson loop in a $U(1)$ factor of the gauge
group can be viewed as a global vortex loop for the associated topological
$U(1)_{J}$ symmetry.

One of our results was the definition of the supersymmetric version of
the vortex loop. This was accomplished by identifying the
fields in the abelian vector multiplet which needed to be turned on
to create the right type of singularity. It turns out that, besides
the singular gauge connection, we needed to also give an $\textit{{imaginary}}$
background value to the auxiliary scalar $D$. This is an interesting
example of the fact that background fields need not satisfy the reality
conditions usually imposed on the dynamical fields of the theory.
We proceeded to evaluate the expectation value of a supersymmetric
defect loop, defined on a great circle on $S^{3}$, using localization.
The result could be anticipated by considering the $SL(2,\mathbb{Z})$
action and, indeed, had already been derived from the 4d perspective.
We have given, by using and comparing two different regularization
methods, an additional microscopic derivation.

The supersymmetric vortex loop plays a central role in mirror
symmetry of 3d gauge theories. This class of dualities has the property
that it exchanges flavor symmetries with the topological symmetries
associated to the abelian factors of the dual gauge group. As a consequence,
the duality exchanges (the supersymmetric versions of) $U(1)$ Wilson
loops with abelian vortex loops. Identifying such entries in the duality
dictionary is an important step towards, possibly, proving the duality
for the full quantum theory. We have demonstrated that the expectation
values for the dual loop operators match, in simple examples, by using
localization and the matrix model.

The analysis presented here has a natural extension to non-abelian
defects. The definition of such an operator can require additional
steps to ensure gauge invariance. For a defect in a global non-abelian
symmetry group the definition is similar to the abelian case and the
results can be read off from Section \ref{sec:Localization-of-the-supersymmetric-defect-operator}
by conjugating the defect data into the Cartan of the flavor group.
When the defect appears in a dynamical gauge field, the localization
procedure for the vector multiplet is modified. The result in the
case of pure Chern-Simons theory is known to require a quantization
of the data entering the definition of the defect, the overall effect
being, again, the insertion of a Wilson loop operator in some representation
\cite{Moore:1989yh}. A naive analysis would imply that this result
is not affected by the presence of additional charged matter. However,
the mapping of such operators under mirror symmetry for non-abelian
theories and under Seiberg-like dualities requires further investigation.

\end{document}